\documentclass[aps,pra,preprint,groupedaddress]{revtex4-1}

\usepackage{graphicx}

\begin{document}


\title{Geographic variation of surface energy partitioning in the climatic mean predicted from the maximum power limit}


\author{Chirag Dhara}
\affiliation{Max-Planck-Institute for Biogeochemistry, Jena, Germany.}

\author{Maik Renner}
\affiliation{Max-Planck-Institute for Biogeochemistry, Jena, Germany.}

\author{Axel Kleidon}
\affiliation{Max-Planck-Institute for Biogeochemistry, Jena, Germany.}


\date{\today}

\begin{abstract}
Convective and radiative cooling are the two principle mechanisms by which the Earth's surface transfers heat into the atmosphere and that shape surface temperature. However, this partitioning is not sufficiently constrained by energy and mass balances alone. We use a simple energy balance model in which convective fluxes and surface temperatures are determined with the additional thermodynamic limit of maximum convective power. We then show that the broad geographic variation of heat fluxes and surface temperatures in the climatological mean compare very well with the ERA-Interim reanalysis over land and ocean.  We also show that the estimates depend considerably on the formulation of longwave radiative transfer and that a spatially uniform offset is related to the assumed cold temperature sink at which the heat engine operates. \end{abstract}


\maketitle

\section{Introduction}
Convective cooling along with longwave radiative cooling are the principle mechanisms by which the Earth's surface reaches steady state with the radiative heating by solar shortwave and downwelling longwave radiation. Convective cooling comprises cooling by  sensible and latent heat fluxes, shaping the surface-atmosphere energy and mass exchange. Latent heat exchange is associated with evaporation, leads to cloud formation and global heat redistribution, hence the relative fraction of radiative to convective cooling at the surface is a critical determinant of surface temperature and also impacts the magnitude of the hydrologic cycle. 

Despite its significance, the detailed physical description of convection and the hydrologic cycle involving the strengths of various associated feedbacks remains one of the pressing challenges to climate modeling \citep{TechSummaryAR5_ipcc, Mauritsen2015}. The problem central to determining the energy partitioning is that the turbulent nature of convection coupled with its small scale organization relative to present day climate model resolutions necessitate empirical convective parametrizations \citep{Stensrud2009}. While high resolution large-eddy simulations provide better estimates, global runs are impractical with present day computational resources \citep{Hohenegger2015}. Several biases result from the uncertainty in convective parametrizations \citep{Hohenegger2009, Bechtold2014} and these have been linked to important GCM deficiencies such as the persistent wide range in equilibrium climate sensitivity \citep{Sherwood2014} and biases in precipitation \citep{Mueller2014}. In short, the fundamental problem in convective parametrization is identifying the most relevant processes and quantifying their coupling \citep{Stevens2013b}.

We approach convection and its role in shaping surface temperature  by asking the extent to which its magnitude is thermodynamically constrained.  To do so, we view the surface-atmosphere system as a heat engine performing work to maintain convection against viscous dissipation. The maximum possible rate that a perfectly non-dissipative heat engine can perform work at  is specified by the Carnot limit. The resulting expression for work generated, \textit{i.e.} power, is proportional to the product of the convective heat flux ($J$) with the difference of the surface ($T_s$) and atmospheric ($T_a$) temperatures ($\Delta T=T_s - T_a$). However, $J$ is tightly linked to the energy balances that shape $\Delta T$, resulting in a fundamental trade-off in which a greater heat flux results in a reduced temperature difference. This trade-off results in a maximum power limit \citep{Kleidon2013a, Kleidon2013b, Kleidon2014}.  With $J = 0$, the temperature difference, $\Delta T$ is greatest, yielding no power.  Yet, with a certain value $J = J_{max}$, $\Delta T$ vanishes thus also yielding no power.  This implies that power has a maximum value for optimal values, $0 \leq J_{opt} \leq J_{max}$ and $\Delta T_{opt}=\Delta T_{opt}(J_{opt})$. This trade-off is critical to our approach and has also been previously used in the context of the proposed principle of Maximum Entropy Production (MEP) \citep{Paltridge1975, Ozawa1997, Lorenz2002, Ozawa2003, Kleidon2005z}. 

We use this limit in conjunction with a simple energy balance model and ask whether the broad climatological patterns and approximate magnitudes of convective heat fluxes and surface temperatures may be estimated from total surface radiative heating (by which we mean here the sum of net shortwave and downwelling longwave radiation absorbed at the surface).  We compare these estimates to a ERA-Interim  climatology \citep{Dee2011}. In doing so, our goal is not to get the most accurate prediction of these fluxes, but rather a climatological estimate of the surface energy balance that is based on physical first principles in a transparent way that complements much more complex climate modeling approaches.

\section{Methods and data sources}
\label{sec:methods}

\begin{figure}
\includegraphics[width=30pc]{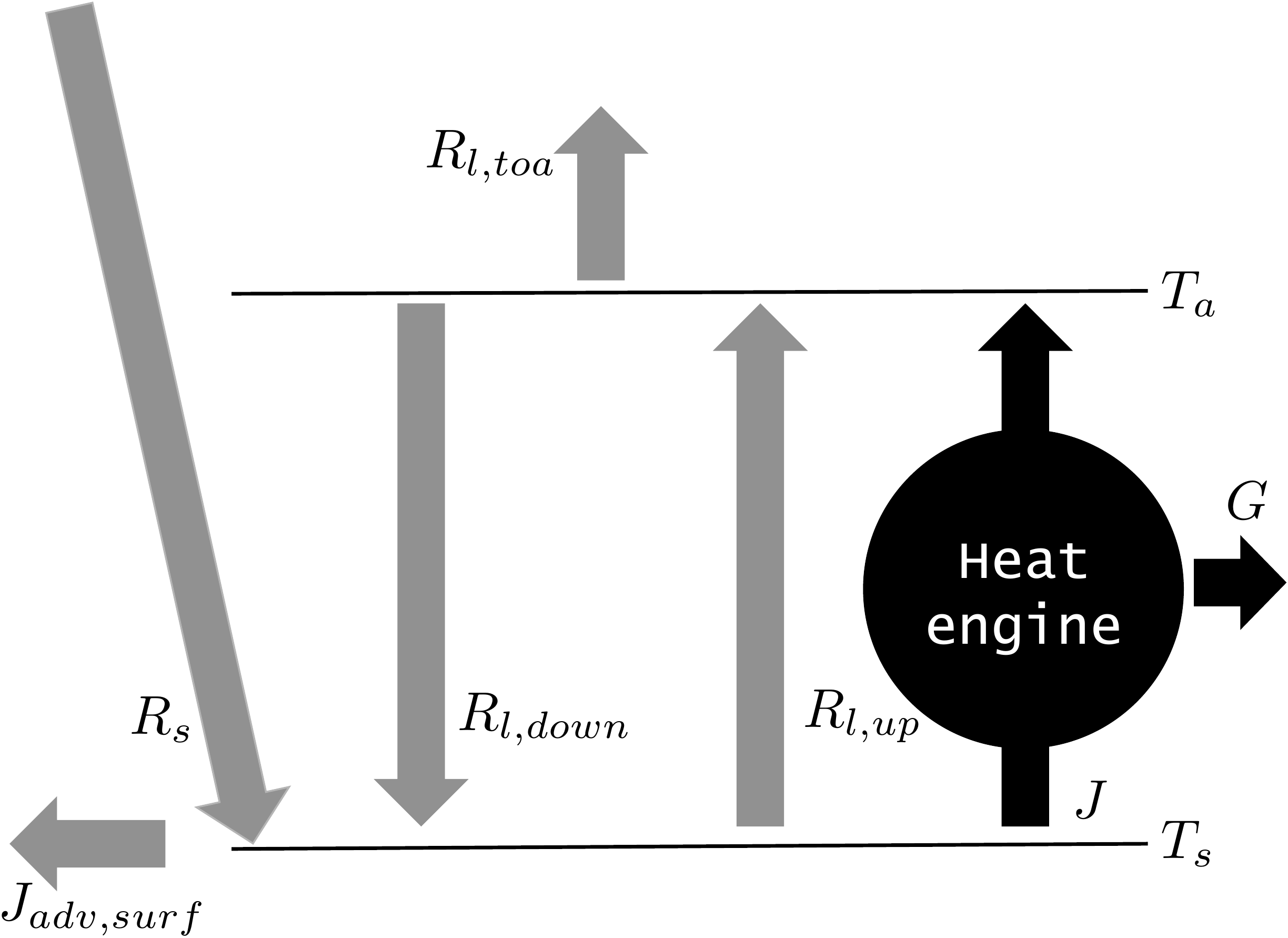}
\caption{The two-box energy balance model with the convective heat engine generating power to sustain atmospheric motion against dissipation. The surface incident net shortwave radiation $R_{s,surf}$, surface longwave forcing $R_{l,down}$, the top of the atmosphere (TOA) longwave flux to space $R_{l,toa}$ and (d) surface advection $J_{adv,surf}$ comprise the forcings while the predicted quantities are convective flux $J$ and surface temperature $T_s=(R_{l,up}/\sigma)^{1/4}$. The convective power $G$ is maximised subject the energy balance constraints.}
\label{heat_engine_new}
\end{figure}

We use a simple two-box energy balance model, with one box representing the surface and the other, the atmosphere (Fig. \ref{heat_engine_new}) following \citep{Kleidon2013a}. The convective exchange between the two boxes is treated as a heat engine.
We infer the maximum power limit for convection from the combination of the Carnot limit with the surface energy balance.  The Carnot limit describes the maximum power that can be derived from a convective heat flux $J$ and is expressed by
\begin{equation}
G(J) = J \; \frac{T_s(J) - T_a}{T_s(J)},
\label{eqn:carnot}
\end{equation}
where $G$ is power, or the generation rate of convective motion, and $T_s$ and $T_a$ are the temperatures of the surface and the atmosphere respectively.  In the following, we use the surface energy balance to express $T_s$ as a function of the convective heat flux to derive the maximum power limit of convection.  This limit is then used in a simple, self-consistent way to predict $T_s$ and $J$ from the surface radiative heating flux of net solar and downwelling longwave radiation minus the surface lateral heat advection (over the ocean).

The surface energy balance is described by 
\begin{equation}
R_{s,surf} + R_{l,down} - J_{adv,surf}  = R_{l,up} + J,
\label{eqn:energy_balance_surface}
\end{equation}
where $R_{s,surf}$ is the net solar shortwave radiation incident at the surface, $R_{l,down}$ the downwelling longwave radiation, $R_{l,up}$ the longwave radiation emitted from the surface, $J= H + \lambda E$ is the total convective flux cooling the surface where $H$ and $\lambda E$ stand for the sensible and latent heat fluxes, and $J_{adv,surf}$ the net heat advected laterally below the surface to describe oceanic heat transport.  The surface temperature $T_s$ is related to the upwelling longwave radiation by $T_s = (R_{l,up}/\sigma)^{1/4}$. Then the left hand side of Eqn. \ref{eqn:energy_balance_surface} involves the effective surface forcing and the right hand side, the surface cooling fluxes and can be used to express $T_s$ in terms of the convective heat flux by
\begin{equation} 
T_s(J) = \Big(\frac{R_{s,surf} + R_{l,down}- J_{adv,surf} - J}{\sigma}\Big)^{1/4}.
\label{eqn:Ts_as_funcn_of_J}
\end{equation} 
We prescribe the atmospheric temperature as the radiative
 temperature by setting $T_a = ( R_{l,toa}/\sigma )^{1/4}$ where $R_{l,toa}$ is the total longwave emission to space. It thus sets a highest temperature limit for the atmospheric box below which the exchange is fully radiative.

\begin{figure}
\centering
\noindent\includegraphics[width=40pc]{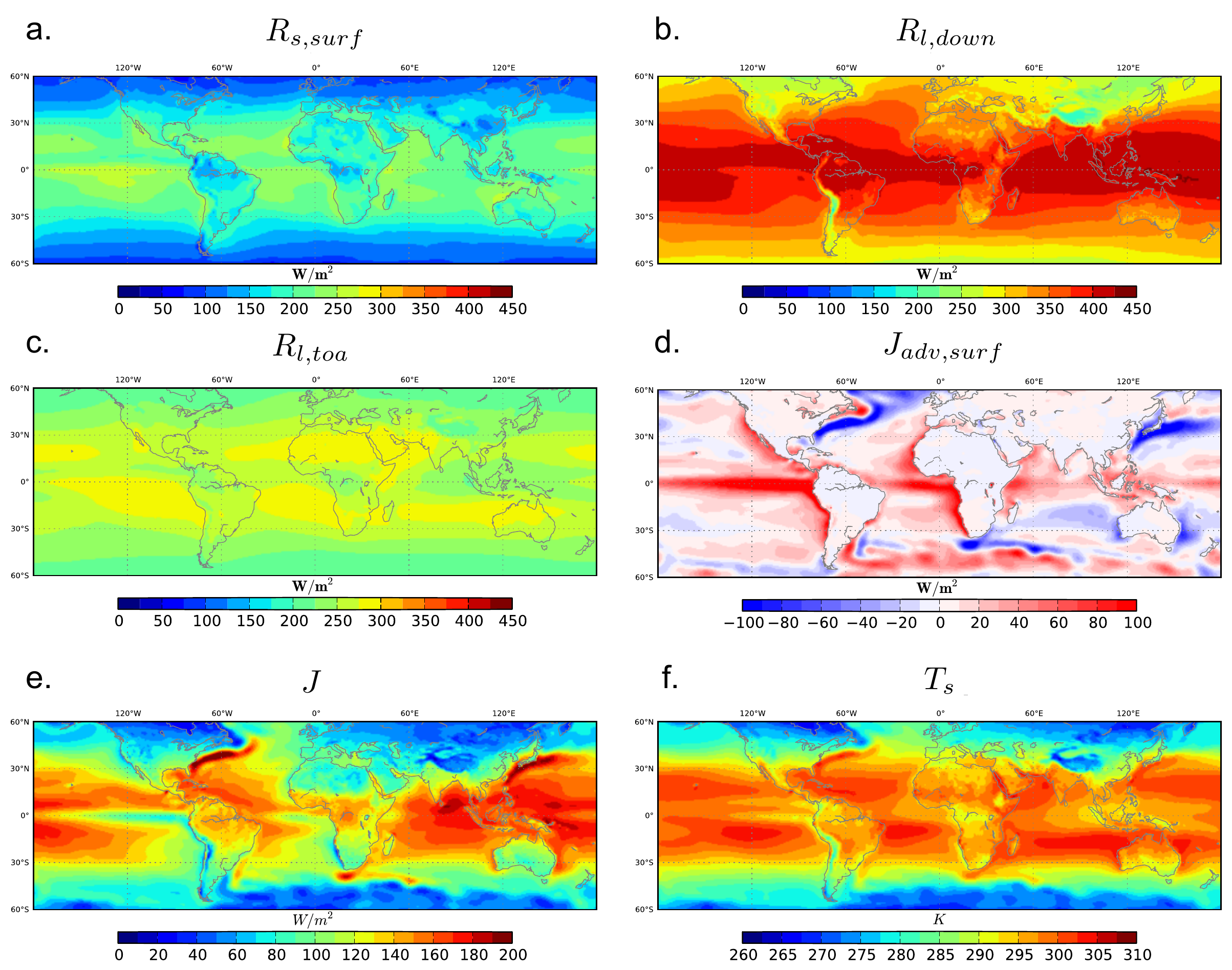} 
\caption{The forcings used in this model and  predictions from the maximum power approach. The forcings are (a) $R_{s,surf}$,  (b)  $R_{l,down}$, (c)  $R_{l,toa}$ and (d) $J_{adv,surf}$. The maximum power predicted  quantities are (e) convective flux $J_{MaxPow}$ and (f) surface temperature $T_{s,MaxPow}$ after modification of the atmospheric temperature using $T_a + 15$ K (see Sec. \ref{sec:results} and Table \ref{Table:bias_J_change_Ta}). }
\label{Fig:forcings_predictions}
\end{figure}

Note that a maximum in power exists for a given forcing described by $R_{s,surf}$, $R_{l,down}$, and $R_{l,toa}$ and a surface advective heat flux $J_{adv,surf}$.  This maximum exists because $G$ increases with $J$ (the first term in Eqn. \ref{eqn:carnot}), but $T_s$ decreases with $J$ (cf. Eqn. \ref{eqn:Ts_as_funcn_of_J}), so that the efficiency decreases with greater $J$ (the second term in Eqn. \ref{eqn:carnot}). The geographic variability of $R_{l,toa}$ is much smaller than those of the surface fluxes (see Fig. \ref{Fig:forcings_predictions}), allowing us to treat $T_a$ as approximately independent of the maximization of power. This leaves $J$ as the  undetermined variable that is diagnosed from the maximization. The expression for power combining the Carnot limit and surface temperature is,
\begin{equation}
G(J) = J \: \Big(1-T_a \:\Big(\frac{R_{s,surf} + R_{l,down}-J_{adv,surf} - J }
{\sigma}\Big)^{-1/4}\Big) .
\label{eqn:combined_eqns_max}
\end{equation}
%
%
%
%
Due to the non-linearity, in the following, we maximise the power numerically to find the optimal solutions $J_{MaxPow}$ and $T_{s,MaxPow} = \Big(\frac{R_{s,surf} + R_{l,down}-J_{adv,surf} - J_{MaxPow} }{\sigma}\Big)^{1/4}$. The solutions to this numeric optimization is used in all the principle results presented in this work. Yet, one of the strengths of simple idealised models is that they often allow analytic solutions (cf \citep{Lorenz2003, Robinson2012}). We find that relaxing the  expression of power in Eqn. \ref{eqn:combined_eqns_max} to a linear function before maximization results in a very simple closed form solution for the convective heat flux. The derivation and the solution can be found in Appendix \ref{Apdx:analytic_J} and Eqn. \ref{eqn:anly_J_lin_G} and prove useful for analytic sensitivity studies such as in Appendix \ref{Apdx:Ta_change_bias}.

We use the ECMWF ERA-Interim reanalysis product \citep{Dee2011} for the forcing data $R_{s,surf}$, $R_{l,down}$, $R_{l, toa}$ to infer $T_a$, and $J_{adv,surf}$ from the imbalance of the surface energy fluxes (see Fig. \ref{Fig:forcings_predictions}). We study geographic variation by performing a grid-by-grid comparison of our results against ERA-Interim data $J_{ecmwf}$ ($= H_{ecmwf} + \lambda E_{ecmwf}$) and $T_{s,ecmwf}$. These data are available at a resolution of $0.75^\circ$ x $0.75^\circ$. To derive climatic means, we computed long term annual means using model output for the period $1991 - 2000$.

\section{Results and discussion}
\label{sec:results}
%
%
%
%
%
%

\begin{figure}
\centering
\noindent\includegraphics[width=40pc]{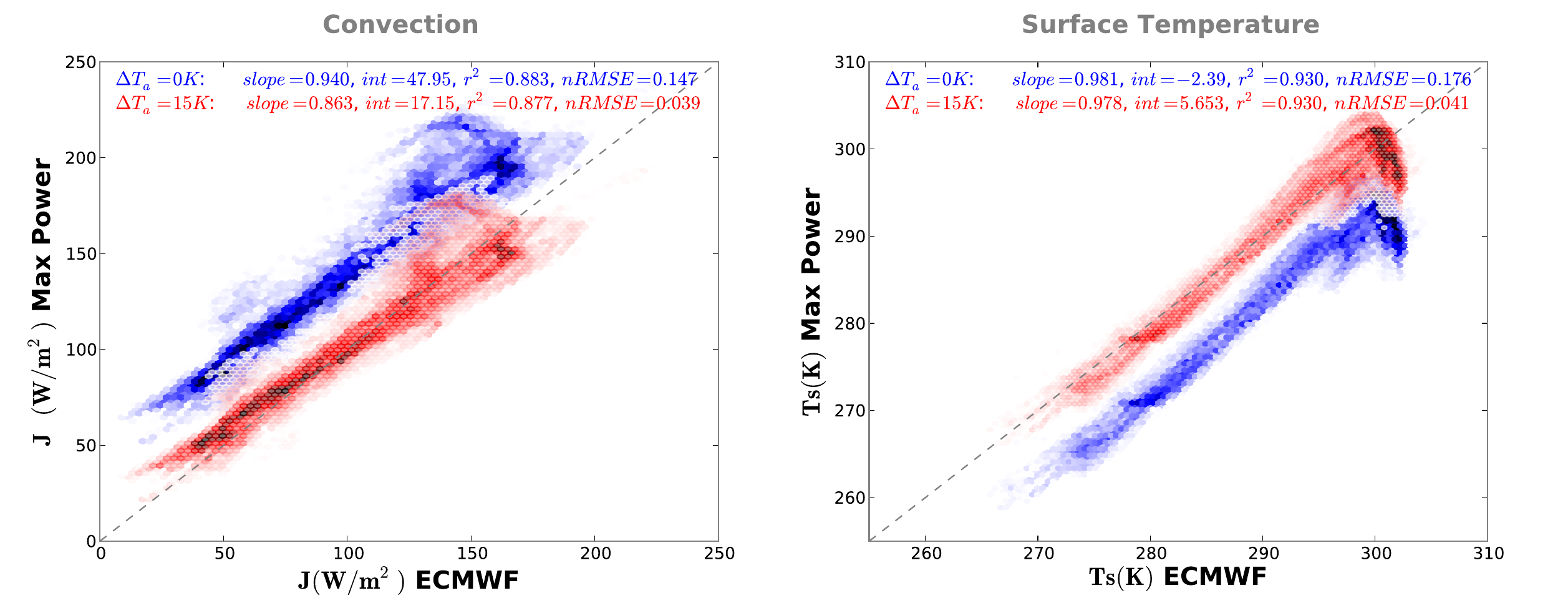} 
\caption{Comparison of the estimates $J_{MaxPow}$  with $J_{ecmwf}$. Slopes  are approximately $1$ in each comparison. Plots in blue are result from using $T_a = (R_{l,toa/\sigma})^{1/4}$ and the ones in red from introducing an offset $T_a' = T_a + \Delta T_a$. Darker colors represent higher density of grid points in these density plots with hexagonal binning. The white hexagons seen in the blue plot are due to low density bins of the overlaying red plot.}
\label{Fig:hxb_J_land_ocean}
\end{figure}

The predicted annual means of the convective heat flux and surface temperature shown in Fig. \ref{Fig:forcings_predictions} (e,f) are compared to ERA-Interim at the grid scale in Fig. \ref{Fig:hxb_J_land_ocean}.  We limit the comparisons to the latitudinal range  $60^\circ$ N to $60^\circ$ S.


The blue points in Fig. \ref{Fig:hxb_J_land_ocean} show the estimates directly obtained by the maximum power limit as described above.  The estimated convective heat flux as well as surface temperature correlate strongly with ERA-Interim ($r^2 = 0.88$ and $0.93$ for $J$ and $T_s$) and show slopes close to one ($= 0.94$ and $0.98$) signifying that our formulation estimates very well the broad geographic variations of both. The analysis is extended to separate  ocean and land in Table \ref{Table:increasing_slope} (scenario $\# 1$) and demonstrates that the geographic variation is captured well in each case.
{\renewcommand{\arraystretch}{1.3}
\renewcommand{\tabcolsep}{0.15cm}
\begin{table}
\centering
\begin{tabular}{c c c c c c c c c}
\hline
\# & Specified & Solution & & Slope & Int. & $r^2$ & nRMSE (\%) & \\
\hline
\multicolumn{1}{c}{}
& &  & Global & $0.94$ & $47.95$ & $0.88$ & $14.7$ \\
1 & $R_{l,down}=$ data & Numeric & Land & $1.06$ & $37.73$ & $0.85$ & $22.9$ \\
& &  & Ocean & $0.92$ & $50.84$ & $0.88$ & $14.6$ \\
\hline
\multicolumn{1}{c}{}
& &   & Global & $0.9$ & $41.9$ & $0.88$ & $11$ \\
2 & $R_{l,down}=$ data & $J=R_{in}\Big(\frac{1}{T_a}\Big(\frac{3}{2}\Big)^{5/4}\Big( \frac{R_{in}}{2\sigma} \Big)^{1/4} -\frac{11}{8}\Big)$  & Land & $1$ & $33.22$ & $0.85$ & $18.1$ \\
& &  & Ocean & $0.88$ & $44.6$ & $0.87$ & $11.5$ \\
\hline
\hline
\multicolumn{1}{c}{}
& &   & Global & $0.51$ & $28.2$ & $0.86$ & $9.9$ \\
3 & $R_{l,net}=k_r (T_s-T_a)$ & $J=R_{s,surf}/2$  & Land & $0.41$ & $43.2$ & $0.51$ & $13$ \\
& &  & Ocean & $0.56$ & $19.2$ & $0.98$ & $9.9$ \\
\hline

\end{tabular}
\caption{Importance of the parameterization of 
longwave fluxes to the geographic variation of convection. The statistics derive from the comparison of $J_{MaxPow}$ to $J_{ecmwf}$ (Fig. \ref{Fig:hxb_J_land_ocean}) for the different scenarios enumerated. The first scenario (row $1$) comprises the comparison of  $J_{MaxPow}$ with $J_{ecmwf}$ using $R_{l,down}$ specified by ERA-Interim data. The second scenario (row $2$) compares the analytic solution of Eqn. \ref{eqn:anly_J_lin_G} with $J_{ecmwf}$. The last  scenario (row $3$) compares results of the simplified two-box model of \citep{Kleidon2013a} that uses a linear approximation for the net longwave radiation $R_{l,net} = k_r (T_s - T_a)$, where $k_r$ may be interpreted as a linearised radiative conductance (for details, see \citep{Kleidon2013a}). The first two rows clearly capture geographic variation significantly better and the magnitude of the correlation $r^2$ is more consistent for both the land and ocean comparisons. The root mean square errors have been normalized by the range of the respective data. Note that biases in $T_s$ are equivalent and opposite to $J$ and analysis need not be performed separately.}
\label{Table:increasing_slope}
\end{table}}
%
%
However, we find that our results are systematically biased, which reflects in the non-zero intercepts ($48$ $W/m^2$ and $-2.4$ $K$) in the respective comparisons. We note that since $J$ and $T_s$ are related through Eqn. \ref{eqn:Ts_as_funcn_of_J}, bias in one results in an equivalent opposite bias in the other as can be seen in Fig. \ref{Fig:hxb_J_land_ocean}.  We attribute this bias to our assumption that the heat engine operates with the radiative temperature inferred from $R_{l,toa}$ as the cold temperature of the heat engine.  It can be shown analytically (see Appendix \ref{Apdx:Ta_change_bias}) that increasing the atmospheric temperature by a global offset $\Delta T_a$ changes the bias while leaving the correlation and slope largely unaffected.  Moreover, the analytic expression can be used together with the observed intercept ($48$ $W/m^2$)  to directly compute the offset minimising the bias and is found to be $(\Delta T_a)_{min} \approx 15.3$ $K$. We perform a numeric sensitivity analysis modifying the atmospheric temperature to $T_a' = T_a + \Delta T_a$ for $\Delta T_a = 0, 5, 10, 15, 20$ $K$. The results of this study (Table \ref{Table:bias_J_change_Ta}) confirm the analytic conclusions showing that the bias indeed  changes uniformly with $\Delta T_a$  and is at a minimum for $\Delta T_a = 15$ K.  The plots in red in Fig. \ref{Fig:hxb_J_land_ocean} include this bias correction, for which the RMSE value is reduced by $\approx 75\%$ for both $J$ and $T_s$.  This suggests  that the heat engine operates with a lower effective temperature difference than $T_s - T_a$, possibly because specifying the atmospheric temperature as the TOA  radiative temperature over-estimates the effective atmospheric depth to which convective heat exchange is sustained. 
{\renewcommand{\arraystretch}{1.3}
\renewcommand{\tabcolsep}{0.15cm}
\begin{table}
\centering
\begin{tabular}{c@{\hskip 0.2in}c@{\hskip 0.2in}c@{\hskip 0.2in}c@{\hskip 0.2in}c}
\hline
$\Delta T_a$ $(K)$ & Slope & Intercept & $r^2$ & nRMSE (\%)   \\
\hline
%
$\mathbf{0}$ & $\mathbf{0.94}$ & $\mathbf{47.95}$ & $\mathbf{0.88}$ & $\mathbf{14.7}$ \\
$5$ & $0.94$ & $47.95$ & $0.88$ & $10.7$ \\
$10$ & $0.89$ & $27.7$ & $0.88$ & $6.8$ \\
$\mathbf{15}$ & $\mathbf{0.86}$ & $\mathbf{17.15}$ & \
$\mathbf{0.88}$ & $\mathbf{3.9}$ \\
$20$ & $0.84$ & $6.21$ & $0.87$ & $5$ \\
\hline
\end{tabular}
\caption{Sensitivity of $J_{MaxPow}$ to $T_a$ i.e. change in bias in the comparison of $J_{MaxPow}$ to $J_{ecmwf}$ by modifying the atmospheric temperature to $T_a' = T_a + \Delta T_a$ for $\Delta T_a = 0, 5, 10, 15, 20$ $K$. The results show that the bias (root mean square error normalised by the range of the data) is lowest for $\Delta T_a = 15K$. $r^2$ remains constant over the range of chosen $\Delta T_a$. The slope reduces by $\approx 10\%$ for $\Delta T_a = 0,15$ $K$ while the error reduces by $\approx 73 \%$. Note that this finding is consistent with the analytic computation (see Eqn. \ref{eqn:J_bias_correction} and the following discussion).}
\label{Table:bias_J_change_Ta}
\end{table}}

We also evaluate the role of the surface longwave heating ($R_{l,down}$) to our results. The results discussed thus far involve specifying $R_{l,down}$ from reanalysis data and the numeric maximization of power. In this formulation, the atmospheric box represents the integrated effects of the atmosphere on the longwave  surface heating. This is the most physically representative case within the idealised $0$D structure of our model.  Row $1$ of Table \ref{Table:increasing_slope} lists statistics of the comparison of $J_{MaxPow}$ and $J_{ecmwf}$. However, as already stated earlier, we also find an  analytic approximation $J_{anly}$ in Appendix \ref{Apdx:analytic_J}. The slope of the comparison of $J_{anly}$ with $J_{ecmwf}$ is globally reduced by approximately $4.5\%$ (row $2$, Table \ref{Table:increasing_slope}) from that of the numeric computation. Thus, the solution reveals the approximate functional dependence of convection on the inputs to the model making it useful for analytic sensitivity studies. 


However, earlier work  \citep{Kleidon2013a} formulated the maximum power approach with an alternative  parameterization. Linearizing the Stefan-Boltzmann law around the reference temperature $T_{ref}$, it was shown that that the 
net longwave radiation can be expressed as $R_{l,net} = k_r (T_s - T_a)$, where $k_r = 4\sigma T_{ref}^3$ represents a "radiative conductance". The strength of these simplifications is that the optimal solutions to maximising power are the simple expressions,  $J=R_{l,net}=R_{s,surf}/2$. In this formulation, the atmospheric box represents not the integrated effect but an unspecified, single level in the atmosphere and cools by emitting uniformly in the upward and downward directions.  Comparisons analoguous to those discussed above are found in row $3$ of Table \ref{Table:increasing_slope} and show that the reduction in slope is considerable being $\approx 40\%$. These approximate solutions were usefully employed to study the surface heat fluxes over land \citep{Kleidon2013c} and the sensitivity of the hydrologic cycle \citep{Kleidon2013b}, however, poorly capture the geographic variability analysed in this work.

We attribute  the significantly greater success of the integrated atmosphere model  to $R_{l,down}$ incorporating the effects of atmospheric radiative transfer and surface forcing induced by  cloud, water vapour and greenhouse gases \citep{Goody1964, Philipona2004}. In other words, the forgoing discussion demonstrates the importance of the magnitude and parameterization of the longwave fluxes to the outcome of the maximum power limit.

Our approach relates closely to the previously proposed principle of maximum entropy production (MEP) \citep{Paltridge1975, Ozawa2003, Kleidon2005z} which states that non-equilibrium thermodynamic systems relax to a state that maximises the rate of entropy production. MEP has previously been used to explain mantle convection \citep{Vanyo1981, Lorenz2002}, meridional heat transport \citep{Paltridge1975, Lorenz2001} and globally averaged atmospheric temperature and convective flux profiles in a 1D model \citep{Ozawa1997}. Maximizing power is equivalent to maximizing entropy production in the two-box model when the power is assumed to be fully dissipated at the cold temperature.  However, the advantage we see of the maximum power formulation is a more intuitive interpretation of the physical limits prescribed by the second law of thermodynamics. Moreover, the magnitude of the predicted power could be tested, for instance, by comparing the power to the generation term of the TKE budget of the convective boundary layer \citep{Moeng1994}.

The highly idealised formulation outlined here is clearly subject to several limitations. The key assumptions we make are: A) steady state, B) the use of the TOA radiative temperature for the cold temperature of the heat engine and C) that the  heat engine operates at the Carnot limit. We now discuss the validity of each of these. We approximate steady states using long-term annual means ($1991-2000$). While this averages out most seasonal and interannual climate variability and heat storage becomes negligible, variations with longer time scales such as the global warming signal may still persist. This may introduce a bias in our comparisons but are expected to be small relative to the magnitudes of the surface fluxes. 

The importance of the assumption (B) is clear from Table \ref{Table:bias_J_change_Ta}. The choice of the atmospheric temperature as the radiative temperature seems appropriate to be used to infer an upper limit on the power generated by the heat engine. Yet, the sensitivity reveals that the effective cold temperature of the heat engine is about 15K warmer than the radiative temperature.  However, it is unclear why the modification at all grid boxes is by approximately the same constant value of $15$ K and leave that as an open question.

Assumption (C) is a strong statement, although it has been used in previous studies \citep{Renno1996, Emanuel1996}, since internal dissipation within the heat engine is expected to result in losses in efficiency of power production. The potential limitations of an idealised heat engine approach have been discussed in later works such as \citep{Pauluis2002a, Pauluis2002} while \citep{Laliberte2015} discuss a general methodology to compute entropy budgets from GCM output and estimate inefficiencies in atmospheric power production. Yet, our argument fundamentally rests on the intuitive coupling between convection and surface temperature defined by the surface energy balance (Eqn. \ref{eqn:energy_balance_surface}) where an increase in one causes a decrease in the other. From this perspective, the tradeoff applies even for efficiencies other than the Carnot limit.

Application of the model at grid scale over the ocean uses advection as an extra forcing (Fig. \ref{Fig:forcings_predictions}) as it cannot be diagnosed within the model. An explicit computation of atmospheric radiative transfer, clouds,  greenhouse and aerosol forcing is not performed here, thus $R_{l,down}$ must be specified from data.

Despite these limitations, we have demonstrated that our minimalist approach captures the geographic variation of  surface energy partitioning based on the local radiative surface heating.

\section{Summary and Conclusions}
Simple $0$D two-box representations have previously been used to study first order responses of the climate to key forcings \citep{Lorenz2010}. Here, we have formulated a simple two-box energy-balance model constrained by ßmaximum power to predict the broad geographic patterns and approximate magnitudes of the convective heat fluxes and surface temperature in the climatological mean.  These magnitudes were predicted from the basic trade-off between radiative and convective cooling of the surface from the prescribed solar radiative forcing, downwelling longwave radiation and lateral heat advection. We also quantified the importance of the  parameterization of downwelling longwave radiation to our results.

Despite several limitations, our results correlate very well with ERA-Interim reanalysis output. Our idealised heat engine approach is distinguished from more complex formulations by the absence of empirical convective parameterization \citep{Stensrud2009}. The surface energy partitioning is instead determined by the hypothesis that the system operates near the thermodynamic maximum power limit.  Our results suggests that convection and surface temperature may indeed be broadly determined by this limit in the climatic mean thus providing useful first-order baseline estimates. This formulation may also serve in future studies as a useful tool - numerically and analytically - to derive limits on the sensitivity surface temperature to changes in surface radiative forcing due to climate change.

\appendix
\section{Approximate analytic expression for the convective flux}
\label{Apdx:analytic_J}

Here, we derive an approximate analytic solution $J_{anly}$ of the maximization of Eqn. \ref{eqn:combined_eqns_max}. 

We linearise this expression around a reference value and maximise it setting $dG(J)/dJ = 0$ to find $J_{anly}$.

\begin{eqnarray}
G(J) &=& J \: \Big[1-T_a \:\Big(\frac{R_{in} - J}{\sigma}\Big)^{-1/4}\Big] \nonumber \\
&=& J \: \Big[1-T_a \:\Big(\frac{R_{in}}{\sigma}\Big)^{-1/4} \Big(1 - \frac{J}{R_{in}}\Big)^{-1/4}\Big]
\end{eqnarray}

The surface energy imposes the constraint $J/R_{in} < 1$. Thus, using the notation $J/R_{in} = x$ the term $(1 - J/R_{in})^{-1/4}$ above can be Taylor expanded around a reference value $x_0$.

The Taylor expansion of the function $f(x) = (1 - x)^{-1/4}$  to first order around the reference $x_0$ is given by,
\begin{eqnarray}
f_{lin}(x) &\approx&  f(x_0) + \frac{df}{dx}\Big|_{x=x_0} (x-x_0)\nonumber \\
&=& (1 - x_0)^{-1/4}\:(1 + \frac{1}{4} (1-x_0)^{-1} (x-x_0))
\end{eqnarray}

With this notation and recalling that $x = J/R_{in}$, the linearised expression of Eqn. \ref{eqn:combined_eqns_max} is,
\begin{eqnarray}
G_{lin}(x)&=& x R_{in} \Big[ 1 - T_a \: \Big(\frac{R_{in}}{\sigma}\Big)^{-1/4} (1 - x_0)^{-1/4}\:\Big( 1 + \frac{1}{4} (1-x_0)^{-1} (x-x_0)\Big) \Big].
\end{eqnarray}

The optimal solution is found by setting $G_{lin}'(x) = 0$ i.e. by setting
\begin{eqnarray}
R_{in} \Big[ 1 - T_a \: \Big(\frac{R_{in}}{\sigma}\Big)^{-1/4} (1 - x_0)^{-1/4}\:\Big( 1 + \frac{1}{4} (1-x_0)^{-1} (2x-x_0) \Big] = 0.
\end{eqnarray}
%
%
%
Using the reference value $x_0 = 1/4$ derived from globally averaged values $J\approx 115$ $W/m^2$ and $R_{in} \approx 160 + 350 = 510$ $W/m^2$, the solution to this equation is,
\begin{eqnarray}
x = \frac{1}{T_a}\Big(\frac{3}{2}\Big)^{5/4}\Big( \frac{R_{in}}{2\sigma} \Big)^{1/4} -\frac{11}{8}.
\end{eqnarray}
%
%
%
Finally, recalling that $x = J/R_{in}$, we find the analytic approximate solution of the maximization of power to be,
\begin{eqnarray}
J_{anly} = R_{in}\,\Big[\frac{1}{T_a}\Big(\frac{3}{2}\Big)^{5/4}\Big( \frac{R_{in}}{2\sigma} \Big)^{1/4} -\frac{11}{8}\Big],
\label{eqn:anly_J_lin_G}
\end{eqnarray}
which is a function of only the effective surface heating $R_{in}=R_{s,surf} + R_{l,down}-J_{adv,surf}$ and the cold temperature of the heat engine, $T_a=(R_{l,toa}/\sigma)^{1/4}$.

\section{Sensitivity of $J_{anly}$ to the prescribed atmospheric temperature $T_a$.}
\label{Apdx:Ta_change_bias}

We show that a specified increase in the atmospheric temperature $T_a' = T_a + \Delta T_a$ results in a spatially uniform offset in convective flux $J$ at the maximum power limit. 

We use the anaytic expression of Eqn. \ref{eqn:anly_J_lin_G},
\begin{equation}
J_{anly} = R_{in}\,\Big[\frac{1}{T_a}\Big(\frac{3}{2}\Big)^{5/4}\Big( \frac{R_{in}}{2\sigma} \Big)^{1/4} -\frac{11}{8}\Big].
\end{equation}

As seen in Fig. \ref{Fig:hxb_J_land_ocean}, this convective flux is uniformly biased with respect to the ECMWF output across grid cells, i.e.,
\begin{equation}
J_{anly} - J_{ecmwf} \approx C_0.
\end{equation}

The expression that results from $T_{a}' = T_{a}+\Delta T_a$ is, 

\begin{eqnarray*}
J_{anly}' &=& R_{in}\,\Big[\frac{1}{T_a+\Delta T_a}\Big(\frac{3}{2}\Big)^{5/4}\Big( \frac{R_{in}}{2\sigma} \Big)^{1/4} -\frac{11}{8}\Big] \nonumber\\
&=& R_{in}\,\Big[ \Big( \frac{1}{T_a} - \frac{\Delta T_a}{T_a (T_a + \Delta T_a)} \Big) \Big(\frac{3}{2}\Big)^{5/4}\Big(\frac{R_{in}}{2\sigma}\Big)^{1/4} -\frac{11}{8}\Big] \nonumber\\
&=& C_0 +J_{ecmwf} - R_{in}\Big[\frac{\Delta T_a}{T_a (T_a + \Delta T_a)} \Big(\frac{3}{2}\Big)^{5/4} \Big(\frac{R_{in}}{2\sigma}\Big)^{1/4} \Big] \nonumber\\
&=& C_0 +J_{ecmwf} - R_{in}\Big[\frac{\Delta T_a}{T_a + \Delta T_a} \Big(\frac{3}{2}\Big)^{5/4}\Big(\frac{R_{in}}{2R_{l,toa}}\Big)^{1/4} \Big].
\end{eqnarray*}
A property of funcions with fractional powers such as $f(x) = x^{1/4}$ is that $\lim f(x) = 1$ for  $x\in (1-\epsilon, 1+\epsilon)$ for $\epsilon < 1/2$. 

In the global climatic mean, $R_{in} = R_{s} + R_{l,down} \approx 510$ $W/m^2$ and $R_{l,toa} \approx 240$ $W/m^2$, hence the term $(R_{in} / 2R_{l,toa})^{1/4}\approx 1.02$. Combining this information with the property of general fractional power functions discussed above, we may approximate $(R_{in} / 2R_{l,toa})^{1/4}\approx 1$ over the range of $R_{in}$ and $R_{l,toa}$ across all grid cells. Using this approximation,
\begin{eqnarray*}
J_{anly}' &=& C_0 +J_{ecmwf} - R_{in}\Big[\frac{\Delta T_a}{T_a + \Delta T_a} \Big(\frac{3}{2}\Big)^{5/4} \Big].
\end{eqnarray*}

For the magnitude of the modification $\Delta T_a = 15 K$ used in Fig. \ref{Fig:hxb_J_land_ocean} and $T_a\approx 255 K$,
\begin{eqnarray}
\label{eqn:J_bias_correction}
J_{anly}' - J_{ecmwf} &=& C_0 - \Big[\frac{\Delta T_a}{T_a + \Delta T_a} \Big(\frac{3}{2}\Big)^{5/4} \Big] R_{in} \\
&\approx& C_0 - 0.09 R_{in}
\end{eqnarray}

Thus, the bias reduction resulting from the shift of atmopheric temperature is relatively uniform despite the large range of $R_{in}$ because of the damping effect of the small constant prefactor attached to it. 

Eqn. \ref{eqn:J_bias_correction} also allows computing the value $(\Delta T_a)_{min}$ which best corrects the bias in $J$. Assigning to $C_0$ the value of the global intercept in scenario $\#1$ of Table \ref{Table:increasing_slope} i.e. $47.95$ $W/m^2$, and setting $C_0 - \Big[\frac{\Delta T_a}{T_a + \Delta T_a} \Big(\frac{3}{2}\Big)^{5/4} \Big] R_{in} = 0$ in Eqn. \ref{eqn:J_bias_correction}, the solution minimising the bias is $(\Delta T_a)_{min} \approx 15.3$ $K$. This is consistent with numeric sensitivity findings in Table \ref{Table:bias_J_change_Ta}.

\begin{acknowledgments}
We thank L. Miller and A. Slamersak for their constructive comments. We also thank the European Centre for Medium-Range Weather Forecasts (ECMWF) for   the reanalysis data used in this work. These may be downloaded from: http://apps.ecmwf.int/datasets/data/interim-full-moda. CD acknowledges the International Max Planck Research School for Global Biogeochemical Cycles (IMPRS-gBGC) for funding. AK and MR acknowledge financial support from the "Catchments As Organized Systems (CAOS)" research group funded by the German Science Foundation (DFG) through grant KL 2168/2-1. 
\end{acknowledgments}

\bibliography{/Users/cdhara/Dropbox/MPI_Papers/library}

\end{document}